\documentclass[onecolumn]{article}
\usepackage[top=1in, bottom=1in, left=1in, right=1in]{geometry}
\usepackage{setspace}
\usepackage{palatino} 
\usepackage{epstopdf}
\usepackage{setspace}

\usepackage[centertags]{amsmath}
\usepackage{amsfonts,amscd,amssymb}
\usepackage{mathtools} 

\usepackage{latexsym}
\usepackage{graphicx}  
\usepackage{epstopdf}
\usepackage{overpic}
\usepackage{xcolor}
\usepackage{rotating}

\graphicspath{{.}}

\usepackage{color}

\title{Lagrangian coherent structures and inertial particle dynamics}
\author{M.~Sudharsan$^1$, Steven L. Brunton$^{1, 2}$, and James J. Riley$^{1,2}$\\
\footnotesize{$^1$ Department of Applied Mathematics, University of Washington, Seattle, WA. 98195-3925}\\
\footnotesize{$^2$ Department of Mechanical Engineering, University of Washington, Seattle, WA. 98195-2420}\\
\vspace{-.3in}}
\date{}
\begin{document}
\maketitle
\begin{abstract}
In this work we investigate the dynamics of inertial particles using finite-time Lyapunov exponents (FTLE). 
 In particular, we characterize the attractor and repeller structures underlying preferential concentration of inertial particles in terms of FTLE fields of the underlying carrier fluid. 
 Inertial particles that are heavier than the ambient fluid (aerosols) attract onto ridges of the negative-time fluid FTLE. 
 This negative-time FTLE ridge becomes a repeller for particles that are lighter than the carrier fluid (bubbles). 
 We also examine the inertial FTLE (iFTLE) determined by the trajectories of inertial particles evolved using the Maxey-Riley equations with non-zero Stokes number and density ratio. 
 Finally, we explore the low-pass filtering effect of Stokes number. 
 These ideas are demonstrated on two-dimensional numerical simulations of the unsteady double gyre flow.  
 \vspace{-.1in}
\end{abstract}

\section{Introduction}
\vspace{-.05in}
Motion of a rigid spherical particle has been a subject of study for almost 200 years, since Poisson analyzed the motion of a pendulum oscillating in air~\cite{poisson1831memoire}. 
He correctly characterized the effect of fluid added mass experienced by a rigid object accelerating through a fluid. 
 Since then, various researchers have worked on accurately predicting the equations of motion of a spherical particle under an uniform flow. 
 In~\cite{Maxey:1983} Maxey and Riley established the equations of motion, the so-called MR equations, and clarified the subject, elucidating the work of past researchers with detailed derivations. 
Although other analytic studies extend the MR equations to more general forms, the results often involve complicated forms hindering their use in repetitive calculations. 
Readers are referred to the comprehensive review article, \cite{michaelides1997review}, for a detailed discussion on MR equations.

Inertial particle flows are abundant both in natural and industrial applications, motivating a deeper understanding of the dynamics of inertial particles. 
 Previous studies have investigated how inertial particles are dispersed by flows, especially by turbulence~\cite{Riley1974pof,squires1991preferential,eaton1994preferential,bec2003fractal,bec2005clustering,bec2007heavy,ravichandran2015caustics}. 
 The inertia of particles has also been studied in the context of gravitational settling velocity and settling time~\cite{stommel1949trajectories,maxey1986gravitational,maxey1987motion,maxey1987gravitational,rubin1995settling}. 
 Preferential concentration of inertial particles in specific regions of a fluid flow has been of particular interest for a number of important applications. 
For instance, inertial particle dispersion by clouds and hurricanes \cite{shaw1998preferential,sapsis2009inertial}, oil spills in the ocean \cite{beron2008oceanic,mezic2010new,nencioli2011surface,olascoaga2012forecasting}, urban pollution dispersion \cite{Tang:2012}, tracking toxic elements \cite{natusch1974toxic}, plankton dynamics in the context of jellyfish predation~\cite{peng:2009}, and capture of inertial particles in aquatic systems~\cite{Espinosa2015jfm} all provide motivation to study the dynamics of inertial particles. 

Our objective is to understand the preferential concentration of inertial particles by applying well established techniques from dynamical systems. 
 Because of the importance of preferential concentration of inertial particles in practical applications~\cite{eaton1994preferential}, we expect this phenomena to gain increasing attention in the following years. 
 In this work, we will investigate the relationship between preferential concentration of inertial particles and the finite-time Lyapunov exponent field (FTLE) of the underlying fluid. 
 We also investigate the inertial FTLE (iFTLE) field derived from measuring the separation of inertial particles in the flow. 
 We extend previous studies like \cite{sapsis2008instabilities,sapsis2009inertial,jacobs2009inertial,haller2008inertial} by simulating trajectories of  particles that are lighter than the ambient fluid (bubbles) in addition to particles that are heavier than the ambient fluid (aerosols). 
 Recently, \cite{Nicolaou2015aps} studied a closely related problem regarding the deposition of aerosol particles in respiratory airways, although they take a different approach, adopting an effective Stokes number to explain deposition efficiency.  
 Here we employ ideas from dynamical system to investigate preferential concentration effects of inertial particles for different density ratios.  
 These inertial particle dynamics are explored in numerical simulations of a physically relevant two-dimensional unsteady double gyre velocity field.

 \subsection{Particle dispersion and preferential concentration}
The dispersion of inertial particles by turbulence is a fundamental topic of current research. 
The dispersion of passive fluid particles is comparatively well understood, providing insights into the spatial structure of turbulent flows. 
 While \cite{taylor1922diffusion} studied the diffusion of heat and other scalar quantities in turbulence, \cite{taylor1954diffusion} focussed on the dispersion of fluid particles by turbulence, sparking considerable interest in the following decades. 
 Two-particle fluid dispersion has become a well-studied quantity used to characterize turbulence, e.g., \cite{Batchelor1952mpcps,fung1998two,malik1999lagrangian,boffetta2002statistics,biferale2005lagrangian,chen2006turbulent}. 
Specifically turbulent-like models are often validated by comparing their two-particle fluid dispersion characteristics with those from direct numerical simulation (DNS)~\cite{malik1999lagrangian}. 
\cite{boffetta2002statistics} investigated Lagrangian relative dispersion in DNS of two-dimensional inverse cascade turbulence; their results agree with Richardson's description of two-particle dispersion. 
 \cite{biferale2005lagrangian}  performed a detailed investigation of particle pair separation in homogeneous turbulence, presenting particle pair separation as a probability density function of separation distance and its second order moment. 
Through the latter they have estimated Richardson's constant, which was found to be in good agreement with the classical theory. 
For an excellent review of two-particle dispersion, see~\cite{salazar2009two}. 
  
 Gravitational settling of particles has been an another motivation to study the dynamics of inertial particles \cite{maxey1986gravitational,maxey1987gravitational,maxey1987motion,rubin1995settling}. 
\cite{maxey1986gravitational}~computed statistics for the motion of small particles settling under gravity in an ensemble of randomly oriented, cellular flow fields that are steady in time. 
They conclude that particles characterized by small free fall velocity and weak inertia show a strong tendency to collect along isolated paths. 
In a sequel, \cite{maxey1987motion} analyzed the trajectories and particle accumulation of bubble and aerosol particles. 
 \cite{maxey1987gravitational} showed for the first time that inertial particles dropped into a Gaussian random velocity field accumulate in regions of high strain rate or low vorticity. 
 This so-called \emph{preferential concentration} was proposed as an explanation for the change in mean settling velocity. 

 Preferential concentration of particles has spurred interest among researchers, especially preferential concentration in turbulent flows \cite{chen2006turbulent,wood2005preferential,shaw1998preferential,fessler1994preferential,eaton1994preferential,squires1991preferential}. 
\cite{squires1991preferential} performed a DNS of isotropic turbulence and showed that particles heavier than the carrier-fluid collect in regions of low vorticity and high strain rate. 
 They concluded that turbulence may actually be inhibiting mixing between inertial particles, because dense particles collect in regions of low vorticity and high strain rate. 
For an excellent review, see \cite{eaton1994preferential}, where preferential concentration is discussed across a wide range of turbulent flows; they point out that preferential concentration is strongest for Stokes numbers near 1. 
Accumulation of inertial particles in turbulent boundary layers has received special attention in some of the recent works. 
\cite{picciotto2005characterization} examined particle accumulation in turbulent boundary layers. 
\cite{guingo2008stochastic} introduced a stochastic Lagrangian model to estimate the particle deposition rate in turbulent flows. 

\begin{figure*}[t!]
\begin{center}
\vspace{-.05in}
\includegraphics[width=\textwidth]{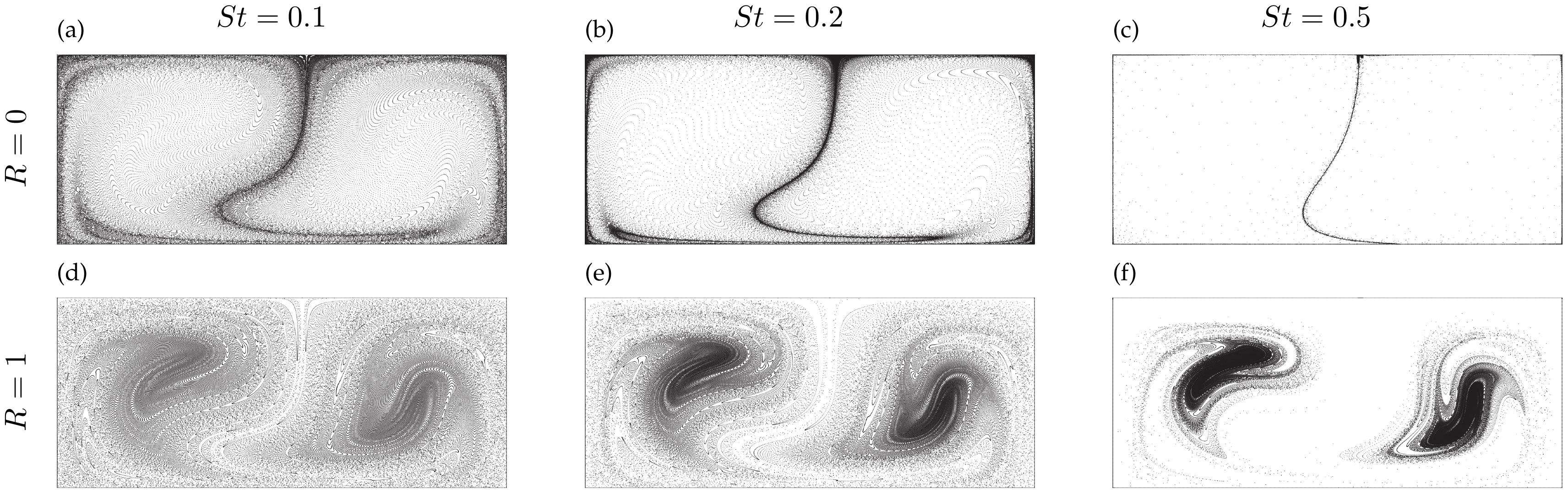}
\vspace{-.2in}
\caption{Simulations with inertial particles uniformly distributed throughout the domain of the double gyre dynamical system, described in Sec.~\ref{Sec:doublegyre}. Preferential concentration patterns formed after advecting the particles for sufficient time are shown.}
\vspace{-.2in}
\label{fig3}
\end{center}
\end{figure*}

We are primarily interested in understanding the patterns formed due to preferential concentration as we vary the Stokes number, $St$, and density ratio, $R$, of particles. 
 The Stokes number represents the non-dimensional particle response time relative to the hydrodynamic time scale of the flow. 
For low $St$, inertial particles behave similarly to fluid particles. 
As $St$ is increased, inertial particles respond more slowly to changes in the flow. 
Figure~\ref{fig3} illustrates the preferential concentration of an initially uniform distribution of particles through the double gyre dynamical system~\cite{Solomon1988,shadden2005definition}, discussed more in Secs.~\ref{Sec:doublegyre} and~\ref{Sec:Results}, for various Stokes numbers and density ratios. 
Aerosol particles ($R = 0$) with a lower Stokes number behave more like the incompressible fluid particles than particles with a higher Stokes number. 
Similar patterns are also observed with increase in $St$ for bubble-like particles ($R = 1$).  Inertial particles are not constrained by incompressibility and have more degrees of freedom. 
However, because of the dissipative nature of the Stokes drag term, the phase space volume of inertial particles decreases monotonically in time. 
As a result we see that with increasing $St$, particles accumulate more rapidly onto thin flow structures.

\subsection{Previous work investigating inertial particles with dynamical systems}
Dynamical systems approaches have been successfully applied to understand the behavior of passive fluid particles. 
 Aref~\cite{Aref:1984} made the critical observation that chaotic particle trajectories may arise from relatively simple velocity fields. 
 Since then, a variety of methods have been used to characterize attractors and repellers in fluids, often based on identifying stable and unstable manifolds that act as separatrices in the flow. 
 In~\cite{rom1990analytical}, separatrices are investigated in the perturbed and unperturbed vortex pair, where perturbation gives rise to heteroclinic chaos. 
 The notion of stable and unstable manifolds for steady systems is analogous to the relatively new extension to general time-varying systems using finite-time Lyapunov exponent (FTLE) fields and Lagrangian coherent structures, which are determined as hyperbolic ridges of FTLE~e.g., \cite{haller2001distinguished,haller2002lagrangian,shadden2005definition,lekien2007lagrangian,green2007detection}. 
 Inertial behavior may also be characterized using tools from dynamical systems~\cite{Babiano:2000}. 
 \cite{rubin1995settling} showed that when inertia is taken as a small parameter the solution to particle motion admits a globally attracting slow manifold. 
 FTLE fields are particularly attractive for analyzing the behavior of inertial particles because they identify regions of attraction or repulsion in the flow. 

Previous studies have investigated inertial particles using FTLE for varying Stokes number for airborne microbes~\cite{Tallapragada:2008,Tallapragada:2011b}, urban flows~\cite{tang2009locating,Tang:2012}, accumulation of inertial particles in the wake of a square cylinder~\cite{jacobs2009inertial}, dissipative transport in the ocean~\cite{beron2014dissipative}, evasive motion of plankton fleeing predators~\cite{peng:2009}, and identifying inertial slow manifolds in hurricane dynamics~\cite{sapsis2009inertial}.   
It is noteworthy to mention the more general treatment of these techniques by Haller and Sapsis~\cite{haller2008inertial}. 
 Their work employed singular perturbation techniques to derive reduced-order equations for the asymptotic motion of small $St$ particles. 
 Integrating the MR equations backwards in time leads to blowup of numerical solutions because of the exponential instability associated with the equations. 
In addition to shedding light on the slow manifold governing the particle motion, their work on the reduced-order equation was also successful in integrating the inertial particles backwards. 
 Therefore the reduced-order equations derived become essential if one is interested in computing backward inertial FTLE.  

Clustering of inertial particles has also been investigated using FTLE for compressible flows~\cite{Perez2015pre}. 
 Modifications to the standard FTLE calculation have also been proposed for inertial particle calculations~\cite{Garaboa2015npg}.

\subsection{Contribution of this work}
In this work, we systematically investigate the dynamics of inertial particles using finite-time Lyapunov exponents.  
In contrast to many of the previous studies, we investigate the Lagrangian behavior of inertial particles for varying density ratios $R$ in addition to varying Stokes number.  
We find that for aerosols ($R<2/3$), inertial particles attract onto FTLE ridges, confirming the results of previous studies; the strength of the attraction onto these ridges is determined by the Stokes number.  
However, for bubbles ($R>2/3$), the FTLE ridges switch from attractors to repellers, so that low-density particles flee from FTLE ridges.  
This provides a dynamical systems perspective on the preferential concentration of inertial particles.  

In addition, we compare the FTLE fields obtained by following passive particles with inertial finite-time Lyapunov exponents (iFTLE) obtained by following inertial particle trajectories.  
We find that fluid FTLE fields provide information about the underlying attractor/repeller structure, while the iFTLE fields provide a measure of inertial particle mixing.  
We also investigate the low-pass filtering effect of Stokes number, demonstrating that particle trajectories, and thus the iFTLE field, attenuate high-frequency oscillations of the fluid velocity field.

\section{Background}
In the present work, we employ finite time Lyapunov exponents in the study of the trajectories of inertial particles. 
Therefore a brief background on computing finite time Lyapunov exponents and on integrating inertial particles trajectories is sketched in this section.
\subsection{Finite time Lyapunov exponents}
Following the formulations of \cite{haller2001distinguished} and \cite{ shadden2005lagrangian}, Lagrangian coherent structures are defined as the locally most attracting or repelling material surfaces in the flow. Identifying these structures is often performed by computing ridges of the finite time Lyapunov exponents~(FTLE), although more recent methods use variational theory~\cite{Farazmand2012chaos}.  

The FTLE is based on the particle flow map, $\Phi_{t_{0}}^{t}$:
\begin{eqnarray}
\Phi_{t_0}^t:{r_{0}} \mapsto  {r}\left( {r_{0}}, t_{0}, t \right)={r}_0+ \int_{t_0}^t{u}({r}(\tau),\tau)\,d\tau,\label{Eq:FTLE}
\end{eqnarray}
where ${u}({r},t)$ is the fluid velocity field, and $r(t)$ is the particle trajectory.  We extract $\lambda({r}_0,t_0,t)$, which is the largest eigenvalue of the Cauchy-Green deformation tensor, $\Delta = \left({D}\Phi_{t_0}^t\right)^*{D}\Phi_{t_0}^t$.  Alternatively, $\sqrt{\lambda}$ is the largest singular value of the flow map Jacobian ${D}\Phi_{t_0}^t$.  The FTLE is then defined as
\begin{equation}
\sigma\left( {r}_0, t_{0}, t \right) = \frac{1}{\left| t-t_{0} \right|} \ln \left( \sqrt{\lambda\left( {r}_0, t_{0}, t \right)} \right).
\label{FTLEeq}
\end{equation}
The FTLE $\sigma$ is a \emph{field}, although it is often computed on a grid of particles.  
Typically, a grid of trajectories are integrated through the dynamics, and the flow map is numerically differentiated to yield the flow map Jacobian.  
 For a more detailed discussion on our formulation, the readers are referred to the excellent book chapter, \cite{shadden2005lagrangian}.

\subsection{Equations for inertial particle dynamics}
In order to track inertial particles numerically, we start with the MR equations~\cite{Maxey:1983}.  Assuming $r(t)$ to represent the position of a particle at time $t$ and $v(t) = \dot{r}(t)$ to represent the corresponding velocity of the particle, then the equation in  dimensional form is the following: 
\begin{equation}
  \begin{aligned}
~~    \mathllap{m_{p} \dot{v}} ~=~&  m_{f} \frac{D}{Dt}u\left( r\left( t \right),t \right) - \frac{1}{2}m_{f} \frac{d}{dt}\left( v-u\left( r\left( t \right),t \right) - \frac{1}{10}a^{2}\nabla^{2} u\left( r\left( t \right),t \right) \right)\\
      & -6\pi a\mu X\left( t \right)+\left( m_{p}-m_{f} \right)g-6\pi a^{2}\mu \int_{0}^{t}{d\tau  \frac{\frac{dX\left( \tau  \right)}{d\tau }}{\sqrt{\pi \nu \left( t-\tau  \right)}}},
\label{MRfulleq}
  \end{aligned}\\
\end{equation}
with\\
\[X\left( t \right) = v\left( t \right) - u\left( r\left( t \right), t \right) - \frac{1}{6}a^{2}\nabla^{2} u \, . \]
Here $m_{p}$ is the mass of the inertial particle, $m_{f}$ the mass of fluid displaced by the particle, $u\left( r\left( t \right),t \right)$ the velocity of fluid at the location $r(t)$ and time $t$, $\mu$ the viscosity of the underlying fluid, $a$ the radius of the particle and $g$ the acceleration due to gravity. The derivative $Du/Dt = \partial u/ \partial t + (u . \nabla) u$, and $d/dt$ is the usual total derivative.
The first term on the right-hand side of Eq.~\eqref{MRfulleq} describes the force exerted by undisturbed fluid on the particle, while the second the term accounts for the added mass effects. Third and fourth terms constitute Stokes drag and buoyancy effects, respectively. The integral term is the Basset history term, and accounts for the effect of particle modifying the flow gradients locally. The $a^{2}\nabla^{2} u$ term is the Faxen correction, performed to justify nonuniform flow effects encountered by the inertial particle.
Equation~\ref{MRfulleq} is valid for small spherical, rigid particles with low particle Reynolds numbers, with Reynolds number computed using the particle radius, $a$, as the length scale, and the slip velocity $|v - u(r(t),t)|$ as velocity. 

Assuming sufficiently small particle radius, $a$, Faxen correction terms can be neglected. The Basset history term can also be neglected assuming that the time interval for a particle to revisit a region it has visited earlier is large in comparison to the time scale of the problem. 
With these assumptions, then non-dimensionalizing Eq.~\eqref{MRfulleq} using the velocity scale, $U$, and the length scale, $L$, of the flow yields:
\begin{equation}
\ddot{r}\left( t \right) = \frac{1}{\mbox{S}t} \left( u\left( r\left( t \right),t \right)-\dot{r}\left( t \right) \right) - W \cdot n + \frac{3}{2}R \frac{d}{dt}u\left( r\left( t \right),t \right),
\label{MRnond}
\end{equation}
Here 
\[\mbox{S}t^{-1} = \frac{6\pi a\mu L}{\left( m_{p}+\frac{1}{2}m_{F} \right)U},     R = \frac{m_{f}}{m_{p}+\frac{1}{2}m_{f}} ,     W  = \frac{m_{p} - m_{f}}{6\pi  a\mu U \mbox{S}t} g,\]
where $n$ is unit vector pointing in the direction of gravity.  For the sake of simplicity, however, gravity will not be considered throughout this work. As a result, Eq.~\eqref{MRnond} can be reduced to the following equation, which contains two non-dimensional quantities which characterize the physical properties of the particle:
 \begin{equation}
 \ddot{r}\left( t \right) = \frac{1}{\mbox{S}t} \left( u\left( r\left( t \right),t \right) - \dot{r}\left( t \right) \right) +  \frac{3}{2}R \frac{d}{dt}u\left( r\left( t \right),t \right).
 \label{MRsimeqns}
 \end{equation}
Here $St$ is the Stokes number, defined as the ratio of the characteristic time of a particle~(or droplet) to a characteristic time of the flow. As $St$ becomes smaller, the particles tend to follow the flow more closely, whereas as $St$ becomes larger the particles tend to deviate from fluid trajectories. $R$ is the density ratio parameter. Notice that with $R = 2/3$ the particles have the same density as that of the carrier-fluid. If $R > 2/3$ then the particles are lighter than the carrier-fluid; we refer to any such particle as a bubble. Similarly particles with $R < 2/3$ are denser than the carrier-fluid and are appropriately called aerosol particles.

 It is noteworthy that particles with same density as that of the carrier-fluid ($R=2/3$) and initiated with $\dot{r}(0) = u(r(0),0)$, track the fluid particles exactly, i.e., $\dot{r}(t) = u(r(t), t)$ and $r(t) = r(0) + \int_{0}^{t} u(r(\tau),\tau) d\tau .$ For a more detailed discussion on the MR equations, the readers are referred to the book chapter, \cite{tel2005chemical}, and to the review, \cite{michaelides1997review}.

\section{Computational methodology}
In order to compute the FTLE field, both for passive or inertial particles, we integrate a set of particle trajectories starting out on a uniform grid from an initial time $t_0$ to a final time $t_f$.  For fluid particles, we integrate particles using the fluid velocity field, as in Eq.~\eqref{Eq:FTLE}. 
If the particles are inertial, we must integrate both the position and velocity of the particle using Eq.~\eqref{MRsimeqns}.  Thus, we may introduce a new state ${q}(t) = \begin{bmatrix} r(t) & v(t)\end{bmatrix}^T$ and integrate as:
\begin{eqnarray}
{q}(t)= {q}(t_0) + \int_{t_0}^{t_f} {f}({q},\tau)\,d\tau,\label{Eq:iFTLE}
\end{eqnarray}
where the dynamics $\dot{{q}} = {f}({q},t)$ from Eq.~\eqref{MRsimeqns} are:
\begin{subequations}
\begin{align}
\frac{d}{dt} r & =  v\\
\frac{d}{dt} v & =  \frac{1}{St}\left(u(r(t),t)-v(t)\right) + \frac{3}{2}R\frac{d}{dt}u(r(t),t).
\end{align}\label{Eq:inertialVF}
\end{subequations}
Here, $r(t)$ represents the position of the inertial particle at time $t$ and $v(t)~=~\dot{r}(t)$ represents the corresponding velocity. 
\label{sec:compmethod}
\subsection{Unsteady vector field: The double gyre}\label{Sec:doublegyre}
To explore the dynamics associated with inertial particles over a range of Stokes numbers and density ratios, we simulate velocity fields and particle trajectories in the well-studied double gyre system~\cite{Solomon1988,shadden2005definition}.  
The double gyre is characterized by the following stream-function:
\begin{equation}
\psi(x,y,t)= A\sin(\pi f(x,t))\sin(\pi y),
 \label{doublegyrevec}
\end{equation}
where
\begin{subequations}
\begin{eqnarray}
f \left( x, t \right) &=& a\left( t \right) x^{2} + b\left( t \right) x,\\
a\left( t \right) &=& \epsilon  \sin  \left( \omega t \right), \\
b\left( t \right) &=& 1-2\epsilon  \sin  \left( \omega t \right).
\end{eqnarray}
\end{subequations}
We consider the double-gyre system defined above over the suitable rectangular domain, [0,~2]~$\times$~[0,~1]. As seen from the equations, the stream function is a combination of sinusoidal composite functions. Specifically the quadratic function $f(x,t)$ leads to periodic oscillations across the domain. Consequently the double gyre system consists of a pair of vortices oscillating back and forth within the cell, [0, 2] $\times$ [0, 1]. In fact, setting $\omega = 0$ results in a steady flow field with a pair of vortices centered at $(0.5, 0.5)$ and $(1.5, 0.5)$. Also, it is noteworthy that any two dimensional planar system defined by a stream function is a Hamiltonian system with its Hamiltonian being the scalar stream function, $ \psi$. Therefore, the phase space volume is conserved in accordance with incompressibility of flows:
\begin{subequations}
\begin{eqnarray}
\dot{x} &=& \frac{\partial \psi}{\partial y},\\
 \label{hamiltonian}
\dot{y} &=& - \frac{\partial \psi}{\partial x}.
\end{eqnarray}
\label{CMethod}
\end{subequations}

In the present work we employed 500~$\times$~250 uniformly spaced particles on a [0,~2]~$\times$~[0,~1] domain. Sufficient accuracy and smoothness are essential to solve the above system numerically since we ultimately want to measure exponential growth in trajectories. 
Therefore care has been taken to employ sufficiently small time stepping with a 4th order Runge-Kutta integrator to integrate particles through the system in Eqs.~\eqref{CMethod}; recall, fluid particles are integrated directly through the velocity field using Eq.~\eqref{Eq:FTLE}, while inertial particles are integrated according to Eq.~\eqref{Eq:iFTLE} using the inertial particle dynamics in Eq.~\eqref{Eq:inertialVF}.  
Therefore care has been taken to employ sufficiently small time stepping with a 4th order Runge-Kutta integrator to integrate the system in Eqs.~(\ref{CMethod}). 
Note that as $St \to 0$, the particles asymptote to fluid trajectories. On account of the nature of the MR equations, these $St \to 0$ particles necessitate relatively shorter time stepping. In order to ensure an accurate simulation of these trajectories, very small time steps were enforced. Once the position of particles are known, the deformation gradient, ${D} \Phi_{t_{0}}^{t}( {r}_{0} )$, can be computed. Specifically we compute the gradient using a finite difference method. Our finite difference method employed central differencing for all the points except on the edges. On the edges we used a single sided difference. FTLE can then be calculated at each location by evaluating the largest singular value of the corresponding deformation gradient matrix. Since FTLE from the above strategy are computed from ${D} \Phi_{t_{0}}^{t}( {r}_{0} )$ and plotted at ${r}_{0}$, new release and tracking of an uniform grid of particles are required for each time instance considered unless an algorithm such as discussed in \cite{brunton2010fast} is employed. In this study, we release and track particles for each time instance considered. For more details on computing FTLE the reader are referred to~\cite{shadden2005lagrangian}.

\section{Results and Discussions}\label{Sec:Results}
Here we investigate the behavior of inertial particle trajectories using techniques from dynamical systems.  
Initially, we explore the preferential concentration of inertial particles for various Stokes numbers and density ratios.  
In this way, we explore the parameter space of the double gyre vector field for a range of Stokes numbers, density ratios, and gyre oscillation frequencies.  
Next, we investigate the finite-time Lyapunov exponent (FTLE) field, both based on the fluid and on the inertial particle trajectories.  
This allows us to establish a connection between the attracting invariant surfaces in the fluid and the preferential concentration of inertial particles.  
Finally, we also explore the low-pass filtering effect as the Stokes number is increased.  

Inertial particles have additional degrees of freedom associated with their momentum, and they are not constrained by incompressibility, even in an incompressible flow.
Because of the dissipative nature of the Stokes drag term, the phase space of inertial particles decreases monotonically in time. 
This is seen in Fig.~\ref{fig3}, where inertial particles contract more aggressively onto thin structures as $St$ is increased.
Additional comparisons of inertial particle phenomena versus their fluid counterparts are reported in~\cite{Babiano:2000}.  Figure~\ref{fig3} also shows that bubbles attract more aggressively onto vortex cores as $St$ is increased.
Note that bubbles appear to be repelled from the regions where aerosols are strongly attracted, and vice versa.  This observation will be revisited in the context of FTLE fields.

\subsection{Finite time Lyapunov exponents}
\begin{figure*}
\begin{center}
\includegraphics[width=\textwidth]{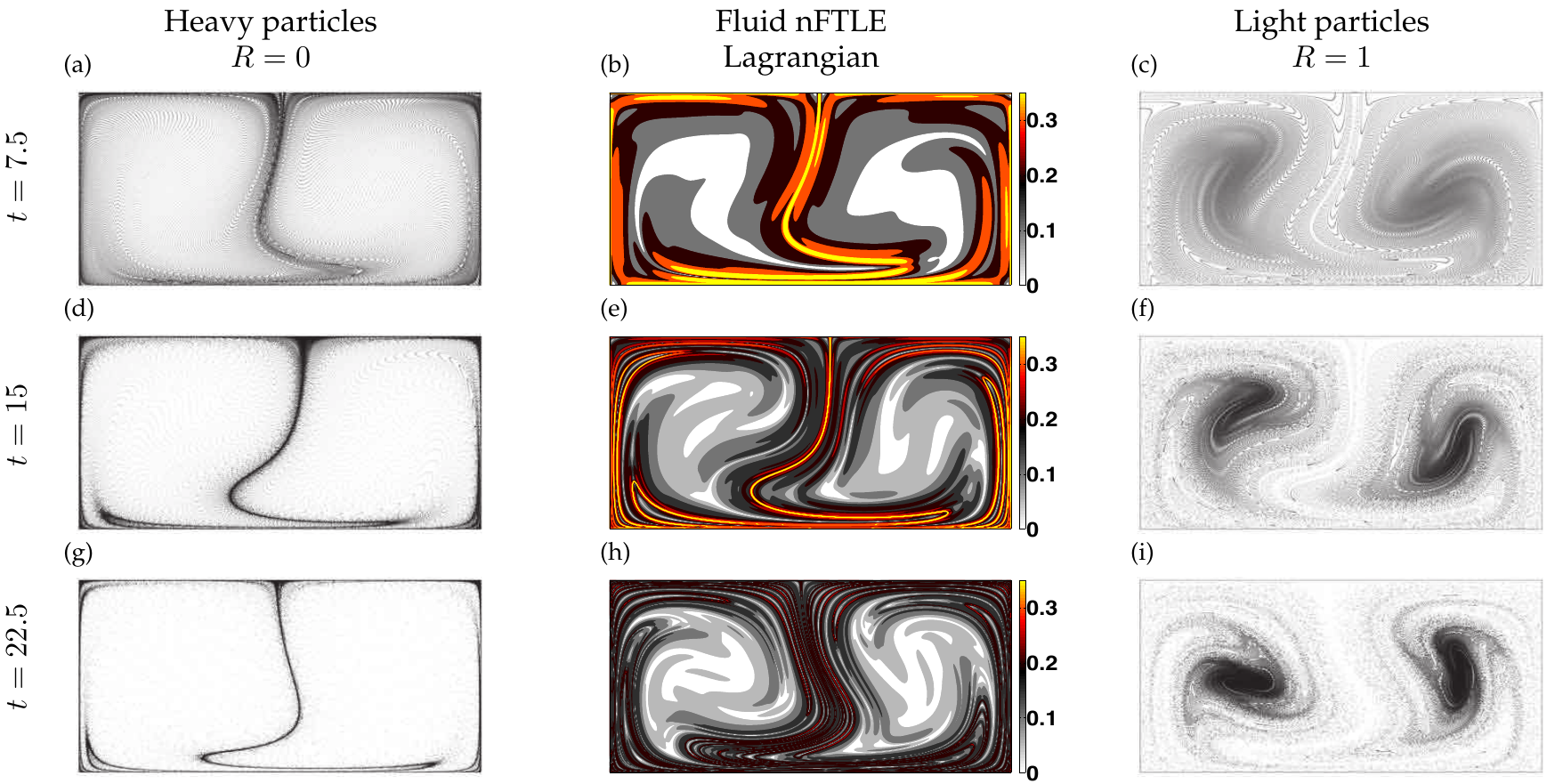}
\vspace{-.1in}
\caption{ Aerosols ($R=0$) \& bubbles ($R=1$) advected using Eq.~\eqref{MRsimeqns} from initial uniform distributions (black patterns). Attracting manifolds of the underlying fluid particles are visualized by computing negative-time Finite time Lyapunov exponents (nFTLE) on flow maps $\Phi_{7.5}^{0}, \Phi_{15}^{0}, \Phi_{22.5}^{0}$ (backward integrated) for (b), (e), (h) correspondingly (red contours). Other parameters are ${\omega={6 \pi}/{10}}$, $St=0.2$, $\epsilon=0.25$, $A=0.1$.}
\label{fig1}
\end{center}
\end{figure*}

Flow visualization is an intuitive and widely employed method to understand fluid mechanics. 
One of the objectives of flow visualization is to reveal coherent features of the flow by initiating passive neutrally buoyant tracer particles at suitable locations. 
It is now known that attracting FTLE ridges, obtained through negative integration time, are highlighted in flow visualization studies. 
Therefore these structures are easier to interpret than repelling FTLE, which are obtained through positive integration time. 
The fact that our data-driven method to evaluate FTLE can identify such complementary coherent features of a flow is one of the significant contributions of FTLE. 
In this section we numerically demonstrate the relationship between attracting FTLE and the preferential concentration of inertial particles.  
Specific distinctions between aerosol-like particles, $R = 0$, and bubble-like particles, $R =1$, are emphasized.

Figure~\ref{fig1}~(b)~(e)~$\&$~(h)~(red contours) numerically visualize the attracting manifolds of fluid particles, and are therefore non-inertial, fluid tracers. 
Furthermore in order to visualize attracting structures, we reverse the direction of integration. 
In other words, negative-time FTLE (nFTLE) were evaluated on non-inertial flow maps integrated backwards in time, $\Phi_{7.5}^{0}, \Phi_{15}^{0}, \Phi_{22.5}^{0}$ as shown. 
Ridges of nFTLE represent exponential stretching of the fluid flow when integrated backwards in time. 
Therefore when fluid particles are advected forward in time, these ridges act as attracting material lines. 
Note that the density of FTLE ridges increases in Fig.~\ref{fig1} for increasing time. 
This indicates increasing mixing between particles initiated at the proximity of the ridges; fluid particles do not leave the cell $[0, 2]$ $\times$ $[0, 1]$.

Figure~\ref{fig1}~(a),~(d),~(g) shows aerosol ($R=0$) particle trajectories at various times. 
Similarly Fig.~\ref{fig1}~(c),~(f),~(i) represent bubble ($R=1$) trajectories advected forward in time. 
It is well-known that inertial particles preferentially concentrate depending on their inertia, i.e., preferential concentration effects provide a rough estimate of particle trajectories based on the local vorticity and the inertia of the particles. 
However, nFTLE of the underlying fluid system pinpoints these zones more accurately. 
Thus, they act as a template for aerosol ($R=0$) attracting manifolds as shown by a striking resemblance between Fig.~\ref{fig1}~(a),~(d),~(g) and Fig.~\ref{fig1}~(b),~(e),~(h) respectively. 
Therefore, material lines identified by FTLE mediate particle trajectories, supplementing the classical preferential concentration theories.  
While aerosol ($R=0$) particles attract onto the attracting manifolds of the fluids, bubble ($R=1$) particles are repelled from the same. 
This can partly be explained by preferential concentration effects. 
However, material lines of nFTLE accurately pinpoint and reveal these structures. 
These results suggest that employing weakly aerosol particles with $R$ slightly less than $2/3$ may enhance flow visualizations of attracting structures. 
Although neutrally buoyant particles are preferred in flow visualization studies for the purpose of not altering the flow, our results indicate that slightly denser particles with $R$ slightly less than $2/3$ adhere onto the underlying attracting FTLE more aggressively than neutral particles.  

\begin{figure*}
\begin{center}
\includegraphics[width=\textwidth]{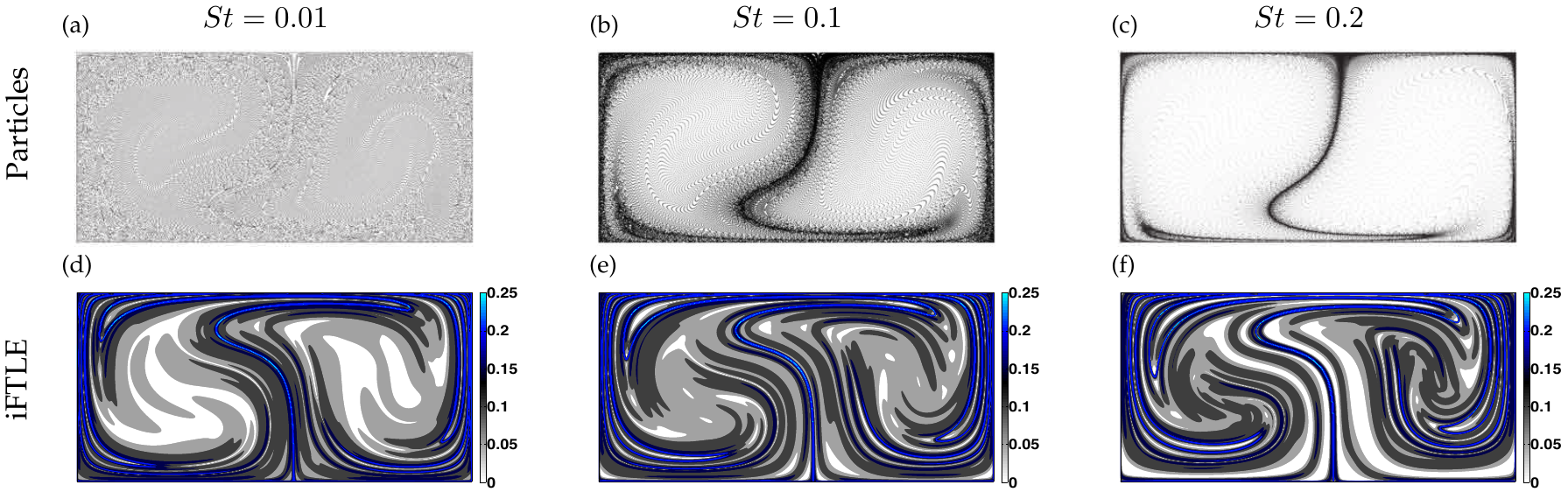}
\vspace{-.1in}
\caption{ Inertial particle trajectories and corresponding positive-time FTLE for various $St$, other parameters are $R=0$, $t=15$, $\omega={6 \pi}/{10}$.}
\label{sfdt}
\end{center}
\end{figure*}

Figure~\ref{sfdt}~(a),~(b)~$\&$~(c) show aerosol ($R=0$) particle positions at $t=15$ for $St$ varying from $0.01$, where particles behave more like passive fluid particles, to $0.2$, where significant dissipation of phase space is characteristic. 
Their corresponding iFTLE are shown in Fig.~\ref{sfdt}~(d),~(e)~$\&$~(f), where inertial particle trajectories are used for the flow map $\Phi$. 
Since these are computed using the respective forward integrated flow maps, $\Phi_{0}^{15}$, the ridges in contour represent material lines of exponential stretching when integrated forward in time. 
In other words, a pair of particles straddled along a ridge of iFTLE are exponentially apart at $t=15$. 
It is noteworthy that particle positions vary  widely with increase in $St$ between Fig.~\ref{sfdt}~(a),~$\&$~(c). 
Despite the different particle trajectories, the iFTLE fields are relatively similar despite Stokes number, as seen in Fig.~\ref{sfdt}~(d),~(e)~$\&$~(f). 
As a consequence, a pair of particles straddled along a ridge in Fig.~\ref{sfdt}~(f) are most likely to be found in different attractors  of Fig.~\ref{sfdt}~(c) at $t=15$, while a similar pair of particles initially straddled along a ridge from Fig.~\ref{sfdt}~(d) are less constrained in their phase space, as seen in Fig.~\ref{sfdt}~(a). 
This also indicates that increasing $St$ has a significant effect on particle trajectories in our specific flow while having very little effect on the organizing structure of the flow. 
Based on these observations it is also possible to devise a strategy to segregate inertial particles by Stokes number. 
For instance higher $St$ particles can be extracted out of the flow near the attractors. 
The fact that nFTLE of fluid particles act as a skeleton of inertial attractors will aid in segregating the particles.
\begin{figure*}
\begin{center}
\includegraphics[width=\textwidth]{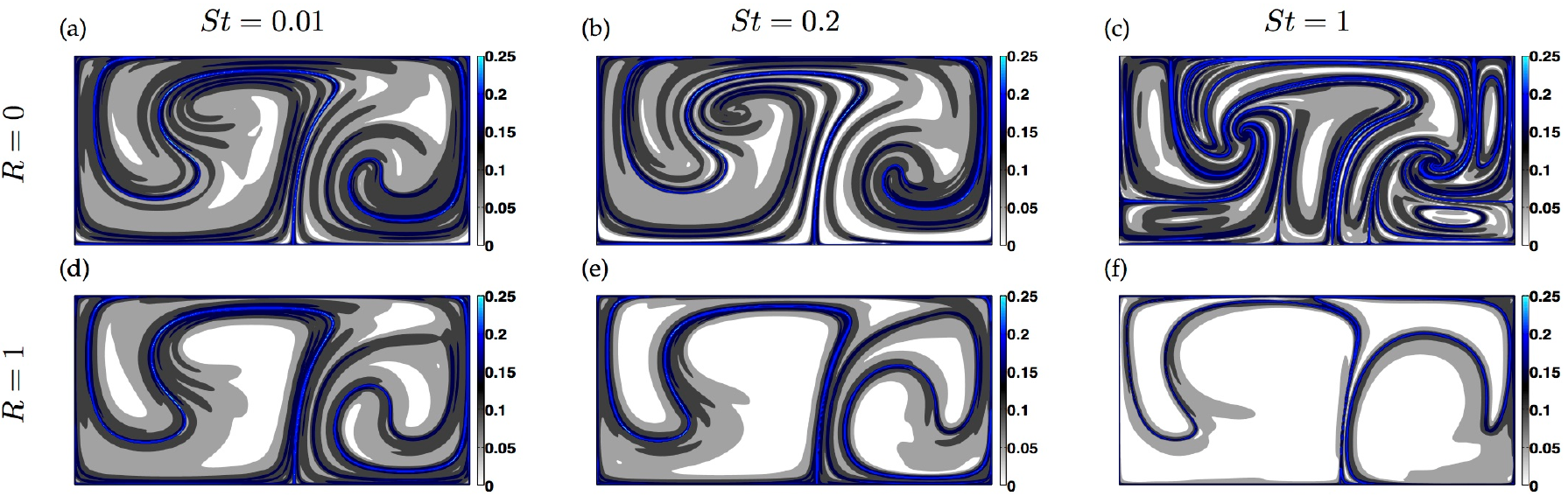}
\vspace{-.1in}
\caption{ Positive-time FTLE for inertial particles with $St$ ranging from 0.01 to 2 $\&$ $R$ from 0 to 1 with $\omega={2 \pi}/{10}$ and $t=15$.}
\label{pFTLEs-mixing}
\end{center}
\end{figure*}

iFTLE of inertial particles with various $St$~and~$R$ are shown in Fig.~\ref{pFTLEs-mixing}. 
Contours of dense aerosol particles, $R=0$, are shown in the first row while the second row represents iFTLE of light particles. 
Increasing $St$ for heavy particles ($R=0$) leads to more ridges as shown. 
On the other hand increasing $St$ for bubbles ($R=1$) lead to relatively fewer ridges of iFTLE; i.e., comparing Fig.~\ref{pFTLEs-mixing}~(c)~$\&$~(g) it is clear that increasing $St$ leads to relatively more and less ridges for aerosols and bubbles, respectively. 
Since iFTLE are measures of exponential stretching of material lines, they can be construed as an indicator of mixing between particles. 
In light of the above, it  is clear that increasing $St$ leads to relatively better mixing for aerosols ($R=0$) while the contrary is true for bubbles. 
This result emphasizes the fact that optimum mixing occurs at different $St$ for bubbles and aerosols.

\subsection{Effect of base flow frequency on small $St$ particles}
\begin{figure*}
\begin{center}
\includegraphics[width=\textwidth]{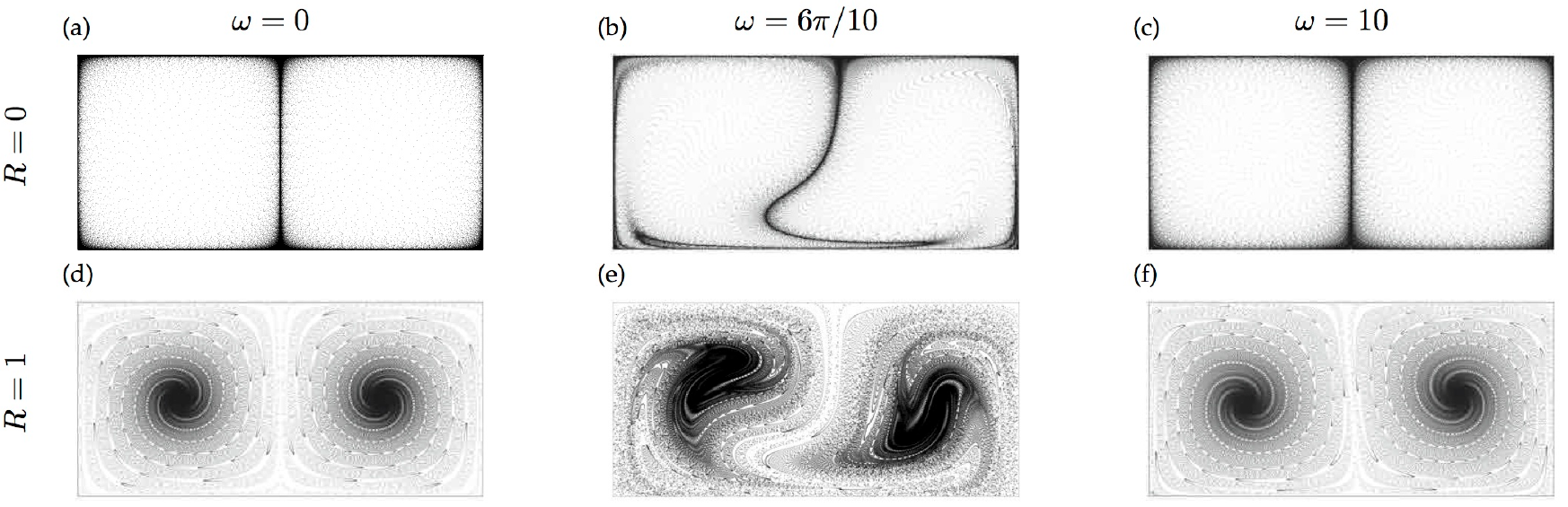}
\vspace{-.1in}
\caption{Inertial trajectories of  particles with $St=0.2$ illustrating the (lowpass filter type) effect of $St$ on the frequency of the flow.}
\label{lowpass}
\end{center}
\end{figure*}

  The double gyre base flow in Eq.~\ref{doublegyrevec} has a characteristic frequency $\omega$ of oscillation.  
 As pointed out earlier, $\omega = 0$ corresponds to steady flow with a pair of steady vortices in $[0, 2]$ $\times$ $[0, 1]$.  
 In Fig.~\ref{lowpass} (a,d), we see that aerosols $(R=0)$ repel from these stationary vortex cores, while bubbles $(R=1)$ attract onto them.  
 As the frequency of oscillation $\omega$ is increased to $6\pi/10$, the particles attract ($R=0$) or repel ($R=1$) from the ridges of fluid FTLE, as seen in (b,e).  
 However, for a larger frequency, $\omega=10$, the inertial particles cannot follow the gyre oscillations because of the low-pass filtering affect of Stokes number.  
 A similar low-pass filtering phenomena is reported in~\cite{bec2006acceleration} for homogeneous, isotropic, fully-developed turbulent flow.  
This work extends these observations to bubbles $(R=1)$ in addition to aerosols $(R=0)$.      
 
\section{Conclusion}
In this work, an attempt to interpret the characteristics of inertial particles through computing finite time Lyapunov exponents (FTLE) has been made. 
Distinctions based on the parameters $St$ and $R$ of the particles and $\omega$, the corresponding frequency of the flow, have been discussed with the help of FTLE fields based on passive fluid particle trajectories as well as inertial FTLE fields based on inertial particle trajectories.  

 Inertial FTLE fields were evaluated on particle trajectories and were used to interpret the dynamics.  
The main result is that inertial particles were observed to attract or repel from the attracting manifolds (nFTLE) of the fluids, depending on whether the particles are aerosol ($R<2/3$) or bubbles ($R>2/3$), respectively. 
This result supplements other well-known results of preferential concentration. 
Our results indicate that an nFTLE of the underlying fluid system acts as a template that organizes the inertial particles. 
Consequently an nFTLE is accurate in pinpointing zones of preferential concentration of inertial particles. 

Increasing the $St$ of particles leads to a higher rate of phase space dissipation. 
However, iFTLE contours of the corresponding system retain most significant ridges as $St$ is increased. 
In addition the magnitudes of the ridges are comparable to each other. 
Therefore a scenario with very similar iFTLE contours with dissimilar particle trajectories is reported. 
Taking into account the attractors formed by higher $St$ particles, the above fact implies that higher $St$ particles straddled along a ridge of iFTLE are most likely to be entrained into different attractors while relatively lower $St$ particles are observed to fill most of the phase space. 
This result has implications for techniques to segregate inertial particles by Stokes number. 
Inertial finite-time Lyapunov exponents are measures of exponential stretching and can be construed as an indicator of the mixing of inertial particles. 
 The fact that increasing $St$ for bubbles results in qualitatively fewer ridges indicates decreased mixing. 
On the other hand increasing $St$ has the opposite effect on aerosols with noticeably more ridges, indicating better mixing. 
The above findings suggest that optimum mixing occurs at different $St$ for bubbles and aerosols.
 
The dynamics of inertial particles were found to be monotonically dissipative as compared to non-inertial flows, where the phase space is preserved. 
We use a periodic stream-function, known as the double gyre, to numerically illustrate this phenomenon. 
The Stokes number, $St$, the non-dimensional particle response time to the flow was increased to demonstrate higher rates of dissipation resulting in increasingly thinner structures. 
These effects were existent irrespective of the density of particles. 
A comparison of these effects on both aerosols and bubbles was drawn by altering $R$, the non-dimensional density parameter. 
Furthermore, increasing $\omega$, the flow frequency, emphasizes the filtering effect of $St$ on particle response. 
Remarkable similarities between the trajectories of particles on steady ($\omega = 0$) and on high frequency ($\omega = 10$) flows have been observed to be in accord with the attenuating effect of $St$ on flow frequency. 
Although such effects have been observed by \cite{bec2006acceleration}, our results extend these observations to bubbles whose densities are less than the base fluid.

The current work relies on a two-dimensional flow based on a planar stream-function to illustrate the use of iFTLE. 
Consequently a natural extension would be to examine these effects on three-dimensional flows. 
This is especially beneficial since most flows in nature and in industrial processes are three dimensional. 
Moreover, our results indicate that increasing $St$ retains the dominant ridges of iFTLE. 
Thus, devising a control strategy to segregate particles based on $St$ is a future possibility. 
Furthermore, based on the fact that aerosol particles are attracted onto the nFTLE of the fluid system, it is suggested that inertial flow visualization can be employed to enhance coherent structure identification in experiments. 

\newpage
\footnotesize{
\bibliographystyle{plain}
\bibliography{references}
}
\end{document}